\documentclass[11pt]{article}

\newcommand{\be}{\begin{equation}}
\newcommand{\ee}{\end{equation}}
\newcommand{\bea}{\begin{eqnarray}}
\newcommand{\nn}{\nonumber}
\newcommand{\eea}{\end{eqnarray}}

\begin{document}

\begin{titlepage}
\begin{flushright}
gr-qc/xxxxxxx\\
\end{flushright}
\begin{centering}
\vspace{.8in}
%


{\large {\bf The effects of curvature correction
 terms on brane cosmology}}
\\

\vspace{.5in} {\bf  E. Papantonopoulos$^{*}$}

\vspace{0.3in}

Department of Physics, National Technical University of Athens,\\
Zografou Campus GR 157 73, Athens, Greece\\
\end{centering}

\vspace{1in}

\begin{abstract} We study the cosmology of the Randall-Sundrum
brane-world where the Einstein-Hilbert action is modified by
curvature correction terms: a four-dimensional scalar curvature
from induced gravity on the brane, and a five-dimensional
Gauss-Bonnet curvature term. The combined effect of these
curvature corrections to the action removes the infinite-density
big bang singularity, although the curvature can still diverge for
some parameter values. A radiation brane undergoes accelerated
expansion near the minimal scale factor, for a range of
parameters. This acceleration is driven by the geometric effects,
without an inflaton field or negative pressures. At late times,
conventional cosmology is recovered.
 \end{abstract}

\begin{flushleft}

 \vspace{1.2in}
 $^{*}$ Invited talk given at Tenth Marcel Grossmann Meeting, Rio de Janeiro,
July 20-26, 2003; lpapa@central.ntua.gr

\end{flushleft}
\end{titlepage}

\section{Introduction}

The Randall-Sundrum~II model \cite{randall} provides a simple
phenomenology for exploring brane-world gravity and associated
ideas from string theory. Matter and gauge interactions are
localized on the brane, while gravity accesses the infinite extra
dimension, but is localized at low energies due to the warping
(curvature) of the extra dimension. The cosmological
generalization of the Randall-Sundrum model is characterized by an
unconventional evolution at early times, while standard cosmology
is recovered at late times  \cite{binetruy,csaki}.

The Randall-Sundrum model is based on the Einstein-Hilbert action
in five dimensions. This gravitational action can be generalized
in various ways. Two important generalizations have been
considered recently. The first is a four-dimensional scalar
curvature term in the brane action. This induced gravity
correction arises because the localized matter fields on the
brane, which couple to bulk gravitons, can generate via quantum
loops a localized four-dimensional world-volume kinetic term for
gravitons \cite{hc,dvali1}. The second is a Gauss-Bonnet
correction to the five-dimensional action. This gives the most
general action with second-order field equations in five
dimensions \cite{lovelock}. Furthermore, in an effective action
approach to string theory, the Gauss-Bonnet term corresponds to
the leading order quantum corrections to gravity, and its presence
guarantees a ghost-free action \cite{zwiebach}.

Here, we investigate the effects of the combined curvature
corrections, from both induced gravity and Gauss-Bonnet. In some
sense, these are the leading-order corrections to the
gravitational action, and there is no obvious way to argue that
one effect is dominant over the other. Indeed, the corrections
operate at different energy levels. Induced gravity introduces
intriguing late-time modifications, which can accelerate the
universe even in the absence of dark energy \cite{deffayet,mmt}.
If the is a brane tension on the brane and a cosmological constant
in the bulk, there are further modifications
\cite{hc,ktt,sahni,mmt} with some interesting astrophysical
implications \cite{astro}. However, one expects that string-theory
type modifications to the Einstein-Hilbert action must also
operate at early times, and so it is sensible to incorporate the
Gauss-Bonnet correction.

At early times, the Randall-Sundrum model gives an unconventional
cosmology, with the Hubble rate $H$ scaling as $\rho$, rather than
$\rho^{1/2}$ as in general relativity. The Gauss-Bonnet correction
to this picture changes the $\rho$ dependence of $H$ to
$\rho^{2/3}$, and therefore an infinite-density big bang is
encountered, as in the Randall-Sundrum case. The combined effect
of Gauss-Bonnet and induced gravity modifications \cite{pap}
eliminates the infinite-density solutions, because the scale
factor is bounded. However, the initial curvature may diverge
since there is a range of parameters for which the solutions start
their evolution with infinite acceleration. In the low-energy
regime of these solutions, the standard cosmology is recovered
(with positive Newton constant).

\section{Friedmann equation on the brane}

The total gravitational action is
 \bea
&& S_{\rm grav}=\frac{1}{2\kappa_{5}^{2}}\int
d^5x\sqrt{-^{(5)\!}g}
\left\{\,^{(5)\!}R-2\Lambda_{5}+\alpha\,\Big[\,^{(5)\!}R^{2}
\right.\nn
\\
&&\left.~{}-4\,^{(5)\!}R_{AB}\,^{(5)\!}
R^{AB}+^{(5)\!}R_{ABCD}\,^{(5)\!}R^{ABCD}\,\Big]\right\}\nn
\\
&&~{} + \frac{r}{2\kappa_{5}^{2}}\int_{y=0} d^4x\sqrt{-^{(4)\!}g}
\left[\,^{(4)\!}R-2\Lambda_{4}\right]\,, \label{action}
 \eea
where the Gauss-Bonnet coupling $\alpha$ has dimensions
$(length)^{2}$ and is defined as
 \bea
\alpha=\frac{1}{ 8g_{s}^{2}}\,,
 \eea
with $g_{s}$ the string energy scale, while the induced-gravity
crossover length scale is
 \bea
r=\frac{\kappa_5^2}{\kappa_4^2}=\frac{M_4^2}{M_5^3}\,.
 \label{distancescale}
 \eea
Here, the fundamental ($M_5$) and the four-dimensional ($M_4$)
Planck masses are given by
 \bea
\kappa_{5}^{2}=8\pi G_{5}=M_{5}^{-3}\,,~~ \kappa_{4}^{2}=8\pi
G_{4}=M_{4}^{-2}\,. \label{planck}
 \eea
The brane tension is given by
 \bea
\lambda={\Lambda_4 \over \kappa_4^2}\,,
 \eea
and is non-negative.

We assume there are no sources in the bulk other than $\Lambda_5$.
Varying Eq.~(\ref{action}) with respect to the bulk metric
$^{(5)\!}g_{AB}$, we obtain the field equations:
 \bea
&& ^{(5)\!}G_{AB} -
\frac{\alpha}{2}\left[\,^{(5)\!}R^{2}-4\,^{(5)\!}R_{CD} \,
^{(5)\!}R^{CD}\right.\nn\\ && \left.~{}
+\,^{(5)\!}R_{CDEF}\, ^{(5)\!}R^{CDEF}\right]\,^{(5)\!}g_{AB}\nn \\
&&~{}+2\alpha\left[\, ^{(5)\!}R \,^{(5)\!}R_{AB}
-2\,^{(5)\!}R_{AC}\,^{(5)\!}R_{B}{}^C\right.\nn\\
&&\left.~{}-2\,^{(5)\!}R_{ACBD}\,^{(5)\!}R^{CD}
+\,^{(5)\!}R_{ACDE}\,^{(5)\!}R_{B}{}^{CDE}\right]\nn \\
&&{}= -\Lambda_{5}\,^{(5)\!}g_{AB}+\kappa_{5}^{2}\,^{\rm
(loc)}T_{AB}\hat{\delta}(y)\,, \label{varying}
 \eea
where $^{(4)\!}g_{AB}=\,^{(5)\!}g_{AB}-n_{A}n_{B}$ is the induced
metric on the hypersurfaces $\{y=$ constant\}, with $n^{A}$ the
normal vector. The localized energy-momentum tensor of the brane
is
 \bea
^{\rm (loc)}T_{AB} \equiv \,^{(4)\!}T_{AB}- \lambda
\,^{(4)\!}g_{AB}-\frac{r}{\kappa_{5}^{2}}\, ^{(4)\!}G_{AB}\,,
\label{tlocal}
 \eea
and we have used the normalized Dirac delta function, $
\hat{\delta}(y)=\sqrt{^{(4)\!}g/\,^{(5)\!}g}\,\,\delta(y)$. The
pure Gauss-Bonnet correction is the case $r=0$, the pure induced
gravity correction is the case $\alpha=0$, and the Randall-Sundrum
case is $r=0=\alpha$.

For a homogeneous and isotropic brane at fixed coordinate position
$y=0$ in the bulk, we get a generic cubic equation in $H^{2}$ \bea
&&{4\over r ^2}\left[1 +\frac{8}{3}\alpha\left(H^2 +{k \over a^2}
+ {\Phi_{0}\over 2} \right) \right]^{2}\left(H^2 +{k \over
a^2}-\Phi_{0}\right) \nn \\ &&~~~{} =\left[H^2 +{k \over a^2}
-\frac{\kappa^{2}_{4}} {3}(\rho+\lambda)\right]^{2}\,,
\label{3fried}
 \eea
where $\Phi_0=\Phi(t,0)$ and $\Phi$ is a solution of the equation
$\Phi+2\alpha \Phi^{2}=\Lambda_{5}/6+\mathcal{C}/a^{4} $ with
$\mathcal{C}$ is an integration constant,
 from
which the Friedmann equations of all known braneworld models can
be derived.

In the limit $r\to 0$, Eq.~(\ref{3fried}) becomes
 \bea
&&\left[1 +\frac{8}{3}\alpha\left(H^2 +{k \over a^2} +
{\Phi_{0}\over 2} \right) \right]^{2}\left(H^2 +{k \over
a^2}-\Phi_{0}\right) \nn
\\ &&~~~{} =\frac{\kappa^{4}_{5}}
{36}(\rho+\lambda)^{2}\,. \label{3gbfried}
 \eea
The single real solution of this cubic which is compatible with
the $\alpha\rightarrow 0$ limit of Eq.~(\ref{3gbfried}), is the
Friedmann equation with Gauss-Bonnet correction~\cite{charmousis}
 \bea
H^{2} +\frac{k}{a^{2} }= \frac{1}{8\alpha}\left(-2+\frac{64I
^{2}}{J}+J\right)\,, \label{friedgb}
 \eea
where the dimensionless quantities $I , J$ are given by
 \bea
&&\!I=\frac{1}{8}(1+4\alpha\Phi_{0})=\pm\frac{1}{8}
\left[1+\frac{4}{3}\alpha\Lambda_{5}+
 \frac{8\alpha \mathcal{C}}{a ^{4}}\right]^{1/2}\!,
 \label{phi1}\\
&&\! J=\!
 \left[
\frac{\kappa^{2}_{5}\sqrt{\alpha}}{\sqrt{2}} (\rho +\lambda) +
\sqrt{{\kappa^{4}_{5}\alpha \over 2} (\rho + \lambda)^{2}
 +(8I )^{3} } \right]^{\!2/3}\!.\label{phi11}
 \eea

In the other limit, $\alpha\to 0$, Eq.~(\ref{3fried}) yields
 \bea
{4\over r^2}\!\left(\!H^2 +{k \over a^2}-\Phi_{0}\!\right)\!
=\!\left[\!H^2 +{k \over a^2} -\frac{\kappa^{2}_{4}}
{3}(\rho+\lambda)\!\right]^{\!2}\,.
 \eea
The solution is the Friedmann equation of the induced gravity
model~\cite{hc,mmt,ktt,sahni}
 \bea
&& H^2+\frac{k}{a ^{2}}=\frac{\kappa_{4}^{2}}{3}(\rho + \lambda)
+\frac{{2}}{r^2}\nn\\&&~{} \pm \frac{1}{\sqrt{3}r}\left[{
4\kappa_{4}^{2}(\rho+\lambda) - 2\Lambda_{5}+{12\over r^2}-
\frac{12\mathcal{C}}{a ^{4}}}\right]^{1/2}.  \label{igr}
 \eea
Finally taking both limits we find the Friedmann equation of the
Randall-Sundrum model.

Returning to the general case of both curvature corrections, we
need the real solution of Eq.~(\ref{3fried}) in the simplest
possible form. We define the dimensionless parameter
 \bea
\beta={ 256\alpha \over 9r^2}\,,
 \eea
and the dimensionless variables
 \bea
P & = & 1+3\beta I\,,\label{parameterP}
\\  Q
&=&\beta\left[\frac{1}{4} +I+\frac{\kappa_{4}^{2}\alpha}{3}
(\rho+\lambda)\right],\label{parameterQ}\\
X& =&\beta\left[\frac{1}{4} + I+\alpha\left( H^2+{k \over
a^2}\right) \right]. \label{parameterX}
 \eea
Then, Eq.~(\ref{3fried}) takes the form
 \bea
X^{3}-P  X^{2}+2 Q X- Q ^{2}=0. \label{x}
 \eea

The single real solution of this equation which is compatible with
the $\alpha\rightarrow 0$ limit of Eq.~(\ref{3fried}), i.e. with
Eq.~(\ref{igr}), is
 \bea
X=\frac{P }{3}-\frac{2}{3}\sqrt{P ^{2}-6 Q } \,\,\cos\left(\Theta
\pm \frac{\pi}{3}\right)\,, \label{indfried}
 \eea
where
 \bea
&&\Theta( P ,Q
)=\frac{1}{3}\arccos\left[\frac{2P^{3}+27Q^{2}-18PQ}
{2(P^{2}-6Q)^{3/2}}\right] \!. \label{omega}
 \eea
This solution corresponds to the positive sign in
Eq.~(\ref{phi1}), while the negative sign does not provide the
correct $\alpha\rightarrow 0$ limit.  The $\pm$ sign
in~Eq.~(\ref{indfried}) is the same as that in~Eq.~(\ref{igr}).
The region in ($ P ,Q $)-space for which Eq.~(\ref{indfried}) is
defined, is
 \bea
&& 1\leq P <{4 \over 3}\,, \label{ineq1}\\ && 2[\,9P -8-(4-3P
)^{3/2}\,]\leq 27 Q \nn
\\&&~~~~{} \leq 3P [\,3-\sqrt{3(3-2P )}\,]\,.\label{ineq2}
 \eea

Finally, we can write the Friedmann equation of the combined
Gauss-Bonnet and induced gravity brane-world as \cite{pap}
 \bea
&& H^{2}+\frac{k}{a ^{2}} =\frac{4-3\beta }{12\beta\alpha }
\nn\\&&~~~{}-\frac{2}{3\beta\alpha } \sqrt{P ^{2}-6 Q
}\,\cos\left(\Theta\pm \frac{\pi}{3}\right)\,. \label{mama}
 \eea

This has a very different structure than its limiting forms,
Eqs.~(\ref{friedgb}) and (\ref{igr}). A closed system of equations
for the brane-world follows if we also consider the continuity
equationm,
 \bea
\dot{\rho}+3H\rho(1+w)=0\,, \label{conservation}
 \eea
where $w=p/\rho\geq -1$ and $\rho\geq 0$. If $w$ is constant, then
$\rho=\rho_0(a_0/a)^{3(1+w)}$, and we can choose $a_0=1$.

\section{Cosmological dynamics}

The dimensionless variable $P $ is a function of $ I $ and carries
the information of the bulk onto the brane, since by
Eq.~(\ref{phi1}) it depends on the bulk cosmological constant
$\Lambda_{5}$ and the mass $\mathcal{C}$ of the bulk black hole.
The dimensionless variable $ Q $ includes information about the
matter and energy content of the brane. These are the key
variables determining the cosmological dynamics.

The four-dimensional scalar curvature term of the induced gravity
and the Gauss-Bonnet term in the five-dimensional space are all
curvature corrections to the Randall-Sundrum model. One could be
led to expect that $r^{2}$ and $\alpha$ are of the same order.
However, this is not necessarily true. The crossover scale $r$ of
the induced gravity appears in loops involving matter particles,
and depending on the mass, it can be arbitrarily large. On the
other hand, the Gauss-Bonnet coupling $\alpha$ arises from
integrating out massive string modes, and depending on the scale
of the theory, it can also be arbitrarily large.

\subsection{No infinite-density big bang}

An important feature arises from inequalities (\ref{ineq1}) and
(\ref{ineq2}), which show that $P$ and hence $Q$ are bounded from
above. Furthermore, Eqs.~(\ref{parameterP}) and (\ref{ineq1}) show
that $I$ is bounded from above (and positive).  Therefore, it
follows from Eq.~(\ref{parameterQ}) that the energy density $\rho$
cannot become infinite, which means that {\it{an infinite-density
singularity $a=0$ is never encountered}}:
 \bea\label{sing}
a(t) \geq a_0>0\,,~~~ \rho(t) \leq \rho_0<\infty\,.
 \eea
This is true independent of the spatial curvature $k$, or the
equation of state.

This result is remarkable since the Gauss-Bonnet correction, which
is expected to dominate at early times, on its own does not remove
the infinite-density singularity~\cite{germani,charmousis,lidnun},
while the induced gravity correction on its own mostly affects the
late-time evolution. However, the combination of these curvature
corrections is effectively ``nonlinear", producing a result that
is not obviously the superposition of their separate effects. In
general terms, the early-universe behaviour is strongly modified
by the effective coupling of the 5D curvature to the
matter~\cite{germani}.

In the pure Gauss-Bonnet theory ($\mathcal{C}\geq 0$), the
early-universe evolves from infinite density at $a=0$. The
Friedmann equation~(\ref{friedgb}) for $\mathcal{C}=0$, or for
$\mathcal{C}>0$, $w>0$, is approximated by
 \bea
H^2+{k\over a^2}\approx
\left(\frac{\kappa^{2}_{5}}{16\alpha}\right)^{2/3}\rho^{2/3}
\,.\label{ll1}
 \eea
For $w>0$, or for $\mathcal{C}=0=k$, the density term dominates
the curvature term, and
 \bea
a\approx \mbox{const}\times\,t^{1/(1+w)}\,.\label{fexp}
 \eea

The Gauss-Bonnet correction causes the universe to expand faster
relative to Einstein gravity, for which
$a\propto\,t^{{2}/{3(1+w)}}$, and to the Randall-Sundrum model,
for which $a\propto\,t^{{1}/{3(1+w)}}$. At the same time, a given
energy density produces a smaller expansion rate in the
Gauss-Bonnet case. This means that there is {\em less} Hubble
friction for a given potential than in general relativity, so that
slow-roll is more difficult to achieve. For the same reason,
scalar perturbations generated during slow-roll inflation will
have a smaller amplitude than those generated at the same energy
density in general relativity. This is opposite to the
Randall-Sundrum model~\cite{mwbh}.

\subsection{Geometric inflation in a radiation universe}

We assume $\mathcal{C}> 0$, i.e. there is a black hole present in
the bulk. Defining the acceleration variable $ f =\ddot{a}/a=\dot
H + H^2$, we obtain from Eqs.~(\ref{mama}) and
(\ref{conservation}) that
 \bea
f&=&\frac{4-3\beta}{12\beta\alpha}+ \frac{\cos(\Theta\pm
\pi/3)}{3\beta\alpha\sqrt{P^{2}-6Q}}
\!\times \nn\\
&&\,{}\times
\Big[c_{1}+\sigma(1-3w)\,|(P-1)^{2}-c_{2}|^{3(1+w)/4}  \nn \\
&& \,{}+(P^{2}-6Q)\,\frac{\dot{\Theta}}{H}\,\tan\left(\Theta\pm
\frac{\pi}{3}\right)\Big]\,, \label{f}
 \eea
where
 \bea &&\!
(P^{2}-6Q)\frac{\dot{\Theta}}{H}= \nn\\
&& = \frac{1}{\sqrt{3}\sqrt{4Q(9P-8)-4P^{2}(P-1)-27Q^{2}}}
\times \nn\\
&&\,{}\times
\Big\{{}\,2\,(2P-9Q)\,\left[\,(P-1)^{2}-c_{2}\,\right]  \nn \\
&&\,\,\,\,\,\,\,\,\,{}-3\sigma(1+w)\,\left[\,3Q-2P(P-1)\,\right]
\times\nn \\
&& \,\,\,\,\,\,\,\,\,\,{}\times |(P-1)^{2}-c_{2}|^{3(1+w)/4}
\,\Big{\}}\,, \label{qz}
 \eea
and
 \bea
c_{1} &=& -2+\beta(3+4\alpha\Lambda_{4})-3\beta^{2}
(3+4\alpha\Lambda_{5})/32\,, \label{Po}\\
c_{2} &=& {3\over 32}\beta^{2}(3+4\alpha\Lambda_{5})\,,\label{Qo}\\
 \sigma &=&
\beta\alpha\kappa_{4}^{2}\rho_{0}\left({8\over 9\beta^{2}\alpha
\mathcal{C}}\right)^{3(1+w)/4}\label{sigma}\,.
 \eea

These equations are formidably complicated, and we do not attempt
an exhaustive analysis. Instead, we show that for a radiation
brane in the presence of a bulk black hole, there is a range of
parameters for which there is inflationary expansion, $f>0$, near
$a_0$.

For a radiation era, $w={1 \over 3}$,
 \bea
f&=&\frac{4-3\beta}{12\beta\alpha}+ \frac{\cos(\Theta\pm
\pi/3)}{3\beta\alpha\sqrt{P^{2}-6Q}} \!\times \nn\\ &&\,{}\times
\Big[c_{1}+(P^{2}-6Q)\,\frac{\dot{\Theta}}{H}\,\tan\left(\Theta\pm
\frac{\pi}{3}\right)\Big]\,, \label{f13}
 \eea
where
 \bea &&\!
(P^{2}-6Q)\frac{\dot{\Theta}}{H}=2\,|(P-1)^{2}-c_{2}|\!\times
\nn\\ &&\,{} \left.\,{}\times \frac{4\sigma
P^{2}-2(2\sigma-1)P-3(2\sigma +3)Q}
{\sqrt{3}\sqrt{4Q(9P-8)-4P^{2}(P-1)-27Q^{2}}} \,.\right.
\label{qz13}
 \eea

We assume that
 \bea\label{inf}
\Lambda_{5}>-{3 \over 4\alpha}\,,
 \eea
and define the additional parameters
 \bea
P_{1} &=& 1+\sqrt{c_{2}\over2}\,,\\
Q_{1}&=& {1\over 12}\left(c_{1}+c_{2}+2+2\sqrt{2c_{2}}\right)\,,\\
\tau &=& Q_{1}-\frac{1}{3}
(P_{1}-1)-\frac{\sigma}{3}(P_{1}-1)^{2}\,. \label{sigm}
 \eea
If the universe expands without limit, $a\rightarrow \infty$,
$t\rightarrow \infty$, then Eq.~(\ref{inf}) is always satisfied,
and $P_{1}, Q_{1}$ are the asymptotic values of $P,Q$. For
$\mathcal{C}> 0$, the variable $P$ plays the role of a time
parameter, since $P(a)$ is monotonically decreasing, with
$P>P_{1}$. Thus, in $(P,Q)$-space, the cosmological evolution is
determined by the curve
 \bea
Q(P)=Q_{1}+\frac{P-P_{1}}{3}+\frac{ \sigma}{3}\left[(P-1)^{2}+
(P_{1}-1)^{2}\right]. \label{curve}
 \eea

There is a well-defined cosmological evolution when this curve
passes through the region defined by the inequalities
(\ref{ineq1}) and (\ref{ineq2}), which in turn depends on the
values of the parameters $P_{1}$, $Q_{1}$ and $\sigma$. A
discussion analogous to the previous one is also valid for
$\mathcal{C}=0$.

One can verify that for $\mathcal{C}\geq 0$ there is a region of
parameter space for which $f$ is positive. The solutions with
$0<f_0<\infty$ represent models that {\em avoid a cosmological
singularity (in density and curvature), and undergo accelerated
expansion from} $a_0$. Furthermore, for $\mathcal{C}>0$, there are
solutions which have infinite acceleration at $a_0$.

The bulk black hole is crucial to the possibility of infinite
acceleration. For $\mathcal{C}=0$, one can show that $f_{0}$
cannot become $+\infty$ for a radiation era. We also note that all
the above results hold independently of the spatial curvature $k$
of the universe.

The accelerating expansion at and near $a_0$, that is driven by
geometric effects, serves as a ``geometric" form of inflation,
very different from conventional scalar field inflation. This
could be interpreted as an alternative to inflaton scenarios,
based on a quantum-gravity correction. However, there remain two
crucial caveats.\\ (1) For $k\geq 0$, there is {\em no exit} from
acceleration for the range of accelerating parameters $\sigma$,
$\tau$ which give infinite $f_{0}$, in the radiation era. This can
be seen, using Eq.~(\ref{mama}), from the fact that the sum of the
first two terms in Eq.~(\ref{f13}) is always positive, since
$P^{2}-6Q$ is monotonically decreasing, and the last term in
Eq.~(\ref{f13}), proportional to $\dot{\Theta}$, is also positive.
The range of $\sigma, \tau$ values gives only a sufficient
condition for acceleration, and we have not been able to
characterize the whole parametric space $(P_{1}, Q_{1}, \sigma,
\beta)$. Therefore, it is still possible that some parameters
exist that lead to an exit from inflation.\\ (2) Those solutions
with $f_0\to\infty$ have {\em a divergence of the Ricci scalar $R$
on the brane, even though the density is finite.} This is
impossible in general relativity or the Randall-Sundrum model,
since in both cases $R=-T$, where $T$ is the trace of the brane
energy momentum tensor. This simple relation breaks down when
there are curvature corrections, and the bulk curvature,
interacting with the brane curvature and matter, plays a decisive
role. Thus, the minimal epoch $a_0$ marks a curvature singularity,
and the brane spacetime geometry breaks down there.

The acceleration-deceleration behaviour of the pure Gauss-Bonnet
and the pure induced gravity models, is very different. For the
Gauss-Bonnet case, we find from Eq.~(\ref{friedgb}) that for
$\mathcal{C}>0$ it is
 \bea
f=-\frac{1}{4\alpha}+
\frac{1}{16\alpha}\left(1-\frac{64I^{2}}{J^{2}}\right)\left(2J+
\frac{\dot{J}}{H}\right)+\frac{16\tilde{c}_{2}}{\alpha J},
\label{fGB}
 \eea
where
 \bea
&& \sqrt{J}\frac{\dot{J}}{H} =
-\frac{2\tilde{\sigma}(1+w)\,(I^{2}-
\tilde{c}_{2})^{{3(1+w)}/{4}}}
{\sqrt{\left[\tilde{c}_{1}+\tilde{\sigma}
(I^{2}-\tilde{c}_{2})^{{3(1+w)}/
{4}}\right]^{2}+(8I)^{3}}}\!\times \nn\\ &&\,{} \left.\,{}\times
\left[ J^{3/2}+{512I \over {\tilde{\sigma}(1+w)}}
(I^{2}-\tilde{c}_{2})^{{(1-3w)}/{4}}\right]
 \!\right. \label{qzGB},
  \eea
with $\tilde{c}_{1}=\sqrt{\alpha}\kappa_{5}^{2}\lambda/\sqrt{2}$,
$\tilde{c}_{2}=(3+4\alpha\Lambda_{5})/192$, and
$\tilde{\sigma}=(\sqrt{\alpha}\kappa_{5}^{2}\rho_{0}/\sqrt{2})
(8/\alpha\mathcal{C})^{3(1+w)/4}$. In $(I,J)$-space, the curve
defining the evolution of the Gauss-Bonnet universe is
 \bea
&&\! J(I)=\Bigg{\{}\tilde{c}_{1}+\tilde{\sigma}\,
(I^{2}-\tilde{c}_{2})^{{3(1+w)}/{4}} \nn\\ &&\,{} \left.\,{}+
\sqrt{\left[\tilde{c}_{1}+\tilde{\sigma}
(I^{2}-\tilde{c}_{2})^{{3(1+w)}/{4}}\right]^{2}+(8I)^{3}}
\,\Bigg{\}}^{2/3} \!\right. \label{curveGB}.
 \eea
Here, $I$ can play the role of time parameter, with $I(a)$
monotonically decreasing for $\mathcal{C}>0$. In the case
$\mathcal{C}=0$, similar expressions are valid. The only candidate
quantity in Eq.~(\ref{fGB}) for producing divergence in $f$ is the
term $2J+\dot{J}/H$. By carefully examining the various
situations, we obtain the result that in the radiation era of the
Gauss-Bonnet universe there is no infinite acceleration. In the
combined theory, because there is no infinite-density regime, the
early universe behaviour cannot be obtained by expanding for large
$\rho$. On the contrary, in the pure Gauss-Bonnet theory, equation
(\ref{ll1}) is the large $\rho$ expansion, from which we can see
furthermore that {\it{there is no initial acceleration, and for
$\rho\rightarrow \infty$, $R\rightarrow +\infty$}}.

Finally, the induced gravity equation~(\ref{igr}) gives in the
radiation era:
 \bea
&& f (\rho)=\frac{\kappa_4^2}{3}(\lambda-\rho)+ \frac{2}{r^2}\nn
\\ &&\,{} \pm {
{\sqrt{2}\over \sqrt{3}\,r}
\left(2\kappa_{4}^{2}\lambda-\Lambda_{5}+{6\over
r^{2}}\right)}\times \nn\\
&&\,\times\left\{2\kappa_{4}^{2}\left[\lambda+\left(1-
{3\,\mathcal{C}\over
\kappa_{4}^{2}\rho_{0}}\right)\rho\right]-\Lambda_{5}+ {6\over
r^{2}}\right\}^{-1/2}\!\!. \label{qrind}
 \eea
We see from Eqs.~(\ref{igr}) and (\ref{qrind}) that among the
solutions of the induced gravity model, there are some which start
with initial singularity $a=0$, as in the conventional model with
$f=-\infty$. Moreover, there is at least one family of solutions
for the branch with the $+$ sign, characterized by the conditions
$2\kappa_{4}^{2}\lambda-\Lambda_{5}+6/r^{2}>0$,
$\kappa_{4}^{2}\rho_{0}<3\mathcal{C}$ and $k\leq 0$, which start
at a finite scale factor with infinite acceleration, qualitatively
similar to our model. Adopting the point of view that the
characteristics of infinite-density avoidance and initial infinite
acceleration are interesting cosmological features which are still
present in the combined induced gravity plus Gauss-Bonnet model,
we can say that the inclusion of the Gauss-Bonnet term has
improved the situation by eliminating all the infinite-density
solutions.

\subsection{Late universe}

For the parameters that allow $a\to\infty$, Eq.~(\ref{mama}) is
approximated as
 \bea
&&\! H^{2}+\frac{k}{a^{2}}\approx
\frac{4-3\beta-\gamma}{12\beta\alpha}+
\nu\kappa_{4}^{2}\rho\,,\label{genlin}
 \eea
neglecting terms $O(\rho^{4/3})$, where the dimensionless
parameters $\gamma$ and $\nu$ are
 \bea
& &\! \gamma=8\sqrt{P_{1}^{2}-6Q_{1}}\,\cos\left(\Theta_{1}\pm
\frac{\pi}{3}\right)\,,\label{gamma}\\ &&\!
\nu=\frac{2}{3\sqrt{P_{1}^{2}-6Q_{1}}}\Bigg[\cos\left(\Theta_{1}\pm
\frac{\pi}{3}\right)+\sin \left(\Theta_{1}\pm \frac{\pi}{3}\right)
\!\times\nn\\ &&\,{} \left.\,\,\,\times
\frac{3Q_{1}+2P_{1}(1-P_{1})}{\sqrt{3}\sqrt{4Q_{1}(9P_{1}-8)+
4P_{1}^{2}(1-P_{1})-27Q_{1}^{2}}}\,\Bigg] \!,\right. \label{nu}
 \eea
with $\Theta_{1}=\Theta(P_{1},Q_{1})$.

First, we observe that the bulk black hole mass $\mathcal{C}$ does
not appear, which means that even if it is non-zero, it decouples
during the cosmological evolution and does not affect the late
universe dynamics. The bulk is felt in the late universe only
through its vacuum energy $\Lambda_{5}$.

Second, for the branch with the $+$ sign in Eq.~(\ref{mama}),
because of the inequalities (\ref{ineq1}), (\ref{ineq2}), it
follows that $\nu>0$. Thus, although the last term in
Eq.~(\ref{mama}) is negative, in the late-time limit it produces
both a negative cosmological constant, $-\gamma$, which
contributes to the total cosmological constant, and a linear
$\rho$ term with {{positive Newton constant}}. For the branch with
the $-$ sign, $\nu$ may be negative. Third, it is seen from the
$a\rightarrow \infty$ limit of Eq.~(\ref{mama}) that the quantity
$4-3\beta-\gamma$ is always non-negative. Therefore, the
conventional cosmology is recovered with positive effective
gravitational and cosmological constants:
 \bea
G_{\rm eff}=3\nu G_{4}\,,~~ \Lambda_{\rm eff}= {4-3\beta-\gamma
\over 4\beta\alpha}\,.
 \eea

In the Gauss-Bonnet case, the late-universe limit of
Eq.~(\ref{friedgb}) is
 \bea
H^{2}+\frac{k}{a^{2}} \approx \frac{\Lambda_{\rm eff}}{3}+{8\pi
G_{\rm eff} \over 3} \,\rho\,,\label{genlinGB}
 \eea
neglecting terms $O(\rho^{4/3})$, where the effective constants
are
 \bea
\Lambda_{\rm eff} & = &
\frac{3}{8\alpha}\left(-2+\frac{64I_{1}^{2}}{J_{1}}+
J_{1}\right),\label{lambdaGB} \\ G_{\rm eff} &=& \frac{G_{5}}
{2\sqrt{2\alpha}}\sqrt{J_{1}}\left[
\frac{J_{1}^{2}-(8I_{1})^{2}}{J_{1}^{3}+(8I_{1})^{3}} \right]
\label{newtonGB}\,.
 \eea
Here $I_{1}, J_{1}$ are the asymptotic values for $a\rightarrow
\infty$ of the variables $I,J$, defined in terms of the parameters
$\tilde{c}_{1}, \tilde{c}_{2}$ of the Gauss-Bonnet model by the
relations $I_{1}=\sqrt{\tilde{c}_{2}}$\,,
$J_{1}=[\,\tilde{c}_{1}+\sqrt{\tilde{c}_{1}^{2}+
(64\tilde{c}_{2})^{3/2}}\,]^{2/3}$\,. The previous remarks
concerning the non-appearance of $\mathcal{C}$ in the above
equations, as well as the positivity of the effective Newton and
cosmological constants, are still valid.

When the brane tension is zero, the Friedmann
equation~(\ref{genlin}) recovers the standard general relativity
behaviour, since the coefficient $\nu$ in Newton's constant
remains positive and nonzero if we set $\lambda=0$. Therefore, if
both curvature corrections are combined, the conventional
cosmology is recovered, even for a tensionless brane. On the
contrary, in the pure Gauss-Bonnet equation~(\ref{genlinGB}), the
brane tension is essential, since $\lambda=0$ implies $G_{\rm
eff}=0$. This is like the pure Randall-Sundrum case, where
positive brane tension is necessary in order to recover the
standard Friedmann equation~\cite{binetruy,sms}.

The late-time limit of the pure induced gravity Friedmann
equation~(\ref{igr}) gives the positive constants
 \bea
&&\! \Lambda_{\rm eff}=\kappa_{4}^{2}\lambda+\frac{6}{r^{2}}\pm
\frac{\sqrt{6}}{r^{2}}\sqrt{\left(2\kappa_{4}^{2}\lambda-
\Lambda_{5} \right)r^{2}+6},\label{lambdaIG}\\ &&\! G_{\rm
eff}=G_{4}\left[1\pm
\left\{{r^2\over6}\left(2\kappa_{4}^{2}\lambda-\Lambda_{5}\right)
+1\right\}^{-1/2}\right]. \label{newtonIG}
 \eea
When there is no brane tension, and even no bulk cosmological
constant, general relativity is still recovered~\cite{dvali1}.

\section{Conclusions}

We studied the cosmology of a brane-world with curvature
corrections to the Randall-Sundrum gravitational action, i.e. a
four-dimensional curvature term of induced gravity and a
five-dimensional Gauss-Bonnet term. The fundamental parameters
appearing in the model are: three energy scales, i.e. the
fundamental Planck mass $M_{5}$, the induced-gravity crossover
energy scale $r^{-1}$, and the Gauss-Bonnet coupling energy scale
$\alpha^{-1/2}$, and two vacuum energies, i.e. the bulk
cosmological constant $\Lambda_{5}$ and the brane tension
$\lambda$. These parameters determine the cosmological evolution
of the brane universe.

We derived the Friedmann equation of the combined curvature
effects, Eq.~(\ref{mama}), which smoothly matches to the induced
gravity equation when the Gauss-Bonnet term vanishes. This
equation has a structure which is quite different from its two
limiting forms. All the solutions of the cosmological model are of
finite density, independently of the spatial curvature of the
universe and the equation of state. This is remarkable, since the
Gauss-Bonnet correction on its own dominates at early times and
does not remove the infinite-density singularity, while the
induced gravity correction on its own mostly affects the late-time
evolution. However, the combination of these curvature corrections
produces an ``interaction" that is not obviously the superposition
of their separate effects. In general terms, the early-universe
behaviour is strongly modified by the effective coupling of the 5D
curvature to the matter.

The late cosmological evolution of our model follows the standard
cosmology, even for zero brane tension, with a positive Newton
constant for one of the two branches of the solutions and positive
cosmological constant.

\section*{Acknowledgments}
This work was done in collaboration with G. Kofinas and R.
Maartens and it is partially supported by NTUA research program
"Thalis".


\end{document}